\documentclass[twocolumn]{article}
\usepackage{dcase2024,amsmath,graphicx,url,times,booktabs,tabularx}
\usepackage{hyperref}
\usepackage{bbm}
\usepackage{multirow}

\usepackage{array}
\usepackage{makecell}
\usepackage{rotating}

\title{Integrating Continuous and Binary Relevances in Audio-Text Relevance Learning \vspace{-6pt}}

\name{Huang Xie, Khazar Khorrami, Okko R\"as\"anen, Tuomas Virtanen \vspace{-6pt}}
\address{Signal Processing Research Center, Tampere University, Finland \vspace{-12pt}}

\begin{document}
    \ninept
    \setlength{\abovedisplayskip}{3pt}
    \setlength{\belowdisplayskip}{3pt}

    \maketitle

    \begin{sloppy}

        \begin{abstract}
            Audio-text relevance learning refers to learning the shared semantic properties of audio samples and textual descriptions.
            The standard approach uses binary relevances derived from pairs of audio samples and their human-provided captions, categorizing each pair as either positive or negative.
            This may result in suboptimal systems due to varying levels of relevance between audio samples and captions.
            In contrast, a recent study used human-assigned relevance ratings, i.e., continuous relevances, for these pairs but did not obtain performance gains in audio-text relevance learning.
            This work introduces a relevance learning method that utilizes both human-assigned continuous relevance ratings and binary relevances using a combination of a listwise ranking objective and a contrastive learning objective.
            Experimental results demonstrate the effectiveness of the proposed method, showing improvements in language-based audio retrieval, a downstream task in audio-text relevance learning.
            In addition, we analyze how properties of the captions or audio clips contribute to the continuous audio-text relevances provided by humans or learned by the machine.
        \end{abstract}


        \section{Introduction}\label{sec:introduction}
        \vspace{-6pt}

        Audio-text relevance learning refers to learning the shared semantic properties of audio samples and textual descriptions.
        It plays an important role in applications such as language-based audio retrieval~\cite{Xie2022Language}.
        Recent studies~\cite{Wu2023Large, Primus2023Advancing} address this problem with similarity learning approaches, which learn intermediate representations of audio samples and texts in a shared embedding space, thereby measuring audio-text relevance by employing a similarity function (e.g., cosine similarity) over these representations.

        Audio-caption datasets (e.g., Clotho~\cite{Drossos2020Clotho}, WavCaps~\cite{Mei2023WavCaps}), which consist of audio samples accompanied by human annotated captions, are widely used for training~\cite{Xie2022Language}.
        Due to the lack of relevance information about audio samples and captions beyond the annotated ones, a binary relevance is adopted between audio samples and captions.
        An audio sample is considered relevant to its annotated caption but irrelevant to all other captions.
        By optimizing a contrastive learning objective (e.g., InfoNCE~\cite{Oord2018Representation}), a learning system is trained to project audio samples and their relevant captions close to each other but far away from the irrelevant ones in the shared embedding space.

        \begin{figure}[!t]
            \centering
            \includegraphics[width=1.0\columnwidth]{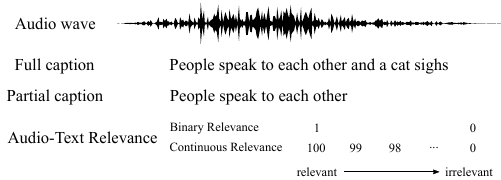}
            \vspace{-24pt}
            \caption{An audio sample with its full caption ``people speak to each other and a cat sighs'', which provides a complete description of its content, and a partial caption ``people speak to each other'', which only describes part of its content. The figure illustrates two categories of audio-text relevance: binary and continuous.}
            \label{fig:introduction_example}
            \vspace{-12pt}
        \end{figure}

        It is likely that audio samples and captions can have varying levels of relevance, ranging from fully relevant to partially relevant.
        For instance, consider an audio sample and its corresponding caption ``people speak to each other and a cat sighs'', as shown in Fig.~\ref{fig:introduction_example}.
        A partial caption ``people speak to each other'' only describes part of the audio sample, as it lacks a description of cat sighs.
        Therefore, it can be seen as partially relevant to the audio sample.
        However, when adopting binary audio-caption relevance, all captions except the one corresponding to the audio sample are regarded as irrelevant.
        In the given example, the caption ``people speak to each other'' will be incorrectly regarded as irrelevant to the audio sample.
        To accurately depict the relevance between audio samples and captions, it is essential to employ non-binary relevance measures (e.g., graded relevance~\cite{Sakai2021Graded, Roitero2021On}).

        Current audio-caption datasets~\cite{Drossos2020Clotho, Mei2023WavCaps} lack annotated non-binary relevance information for their audio samples and captions.
        Our previous study~\cite{Xie2023Crowdsourcing} collected continuous audio-caption relevances for a small subset of Clotho~\cite{Drossos2020Clotho} via crowdsourced subjective assessments.
        Human annotators were asked to assign relevance ratings (ranging from 0 to 100) to audio samples with respect to a given caption.
        It was shown that reducing these ratings to binary relevances for training did not improve model performance on downstream tasks (e.g., language-based audio retrieval)~\cite{Xie2023Crowdsourcing}.
        Conversely, obtaining continuous relevances through subjective assessments is often expensive, being labor-intensive and time-consuming, while training a system typically requires a large amount of data.

        This work proposes an audio-text relevance learning method that leverages both continuous and binary relevances.
        We train modality-specific encoders to project audio samples and texts into a shared embedding space, learning audio-text relevance by computing the cosine similarity of their embeddings.
        During training, we jointly optimize a listwise ranking objective (e.g., ListNet~\cite{Cao2007Learning}) with human-assigned continuous relevance ratings and a contrastive learning objective (e.g., InfoNCE~\cite{Oord2018Representation}) with binary audio-text relevances.
        Experimental results demonstrate the effectiveness of the proposed method, showing improvements in language-based audio retrieval, a downstream task in audio-text relevance learning.
        Additionally, we analyze how properties of the captions or audio clips contribute to the continuous audio-text relevances provided by humans or learned by the machine.

        \vspace{-6pt}

        \section{Proposed Method}\label{sec:proposed-method}
        \vspace{-6pt}

        This section presents the proposed audio-text relevance learning method with continuous and binary relevances.

        \subsection{Audio-Text Relevance Learning}\label{subsec:audio-text-relevance-learning}

        \begin{figure*}[!t]
            \centering
            \includegraphics[width=1.0\textwidth]{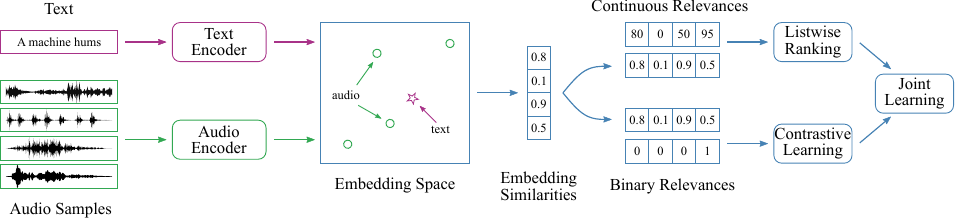}
            \vspace{-24pt}
            \caption{A model-agnostic dual-encoder framework for audio-text relevance learning with continuous and binary relevances.}
            \label{fig:proposed_method}
            \vspace{-12pt}
        \end{figure*}

        Fig.~\ref{fig:proposed_method} presents a model-agnostic dual-encoder framework for audio-text relevance learning.
        Audio samples and texts are projected into a shared embedding space via modality-specific encoders (e.g., audio and text encoders).
        The relevance between audio samples and texts is measured by computing the similarity between their embeddings, such as cosine similarity.
        When training with continuous relevances (e.g., relevance ratings), a listwise ranking objective (e.g., ListNet~\cite{Cao2007Learning}) is computed based on these relevances and the audio-text embedding similarities.
        For binary relevances, a contrastive learning objective (e.g., InfoNCE~\cite{Oord2018Representation}) is calculated from the binary relevances and embedding similarities.
        Using both relevances for training involves a joint learning objective combining the listwise ranking and contrastive learning objectives.

        \subsection{Learning Objectives}\label{subsec:learning-objectives}

        \textbf{ListNet Loss with Continuous Relevances}.
        Given $N$ audio samples $\left\{ y_{1},\cdots,y_{N} \right\}$ and a caption $x$, let $s_{i}$ ($0 \leq s_{i} \leq 100$) be the relevance rating of $y_{i}$ to $x$, for $i=1,\cdots,N$.
        Suppose that $L$ is an ideal ranked list (which is unknown), where $y_{1},\cdots,y_{N}$ are arranged in descending order according to their relevance to $x$.
        The top-one ranking probability of $y_{i}$, denoted by $p(y_{i})$, represents the probability of it being ranked at the top of $L$, given the relevance ratings $\left\{ s_{1},\cdots,s_{N} \right\}$.

        Inspired by~\cite{Cao2007Learning}, $p(y_{i})$ is written as
        \begin{equation}
            \label{eq:ranking_probability_p}
            p(y_{i}) = \dfrac{\phi(s_{i})}{\sum_{j=1}^{N} \phi(s_{j})},
        \end{equation}
        with $\phi(s_{i})$ being
        \begin{equation}
            \label{eq:rating_phi}
            \phi(s_{i}) = \dfrac{s_{i}}{\log_{2} \left( r(s_{i}) + 1 \right)},
        \end{equation}
        where $r(s_{i})$ represents the position of $s_{i}$ in the descending ranked list of $s_{1},\cdots,s_{N}$.
        The top-one probabilities $\left\{ p(y_{1}),\cdots,p(y_{N}) \right\}$ define a probability distribution $P$ over audio samples $y_{1},\cdots,y_{N}$ with respect to the caption $x$.

        In this work, the dual-encoder framework outputs cosine similarity scores between audio and text embeddings as a measure of audio-text relevance.
        Suppose that $\mathbf{a}_{i}$ is the audio embedding of $y_{i}$, and $\mathbf{c}$ is the text embedding of $x$; their similarity score is denoted by $t(\mathbf{a}_{i}, \mathbf{c})$, with $t(\mathbf{a}_{i}, \mathbf{c}) \in \left[ -1, 1 \right]$.

        Similar to~\eqref{eq:ranking_probability_p}, we calculate another top-one ranking probability of $y_{i}$, denoted by $q(y_{i})$, based on the similarity scores $\left\{ t(\mathbf{a}_{1}, \mathbf{c}),\cdots,t(\mathbf{a}_{N}, \mathbf{c}) \right\}$.
        Specifically, the probability $q(y_{i})$ is written as
        \begin{equation}
            \label{eq:ranking_probability_q}
            q(y_{i}) = \dfrac{\exp (t(\mathbf{a}_{i}, \mathbf{c}) / \omega)}{\sum_{j=1}^{N} \exp (t(\mathbf{a}_{j}, \mathbf{c}) / \omega)}.
        \end{equation}
        with $\omega$ being a hyperparameter.
        These top-one probabilities $\left\{ q(y_{1}),\cdots,q(y_{N}) \right\}$ define another probability distribution $Q$ over audio samples $y_{1},\cdots,y_{N}$ with respect to the caption $x$.

        Finally, the ListNet loss is calculated as the cross entropy between the two probability distributions $P$ and $Q$, written as
        \begin{equation}
            \label{eq:listnet_loss}
            L_{\text{ListNet}} = - \sum_{j=1}^{N} p(y_{j}) \log q(y_{j}).
        \end{equation}
        By minimizing~\eqref{eq:listnet_loss}, the dual-encoder framework are optimized for ranking audio samples by their relevance to a given caption.

        \textbf{InfoNCE Loss with Binary Relevances}.
        For the case of binary relevances, InfoNCE loss~\cite{Oord2018Representation} is used for training.
        In InfoNCE, audio samples and captions are considered relevant only if they correspond to each other, and otherwise, they are irrelevant.
        Similar to~\cite{Wu2023Large, Primus2023Advancing}, we use symmetric InfoNCE that tries to classify audio samples as either relevant or irrelevant to a given caption, and vice versa.
        The total InfoNCE loss is then calculated as the sum of two categorical cross entropies of the two tasks.

        \textbf{Joint Loss}.
        To use both continuous and binary audio-text relevances for training, the ListNet~\eqref{eq:listnet_loss} and InfoNCE losses are combined into a joint loss.
        Specifically, the joint loss is written as
        \begin{equation}
            \label{eq:joint_loss}
            L_{\text{joint}} = \alpha \cdot L_{\text{InfoNCE}} + (1 - \alpha) \cdot L_{\text{ListNet}},
        \end{equation}
        where $\alpha$ is a hyperparameter that is chosen from $\left( 0, 1 \right)$.

        \vspace{-6pt}

        \section{Experiments}\label{sec:experiments}
        \vspace{-6pt}

        The proposed method was validated on language-based audio retrieval, a downstream task in audio-text relevance learning.

        \subsection{Audio and Text Data}\label{subsec:audio-and-text-data}

        \textbf{Clotho}.
        All audio samples and texts used in the experiment were selected from Clotho~\cite{Drossos2020Clotho}, which consists of 5,929 audio samples, each with five human annotated captions.
        Clotho is partitioned into three subsets: a development set with 3,839 audio samples, a validation set with 1,045 audio samples, and an evaluation set with 1,045 audio samples.

        \textbf{Continuous Relevances}.
        Our previous study~\cite{Xie2023Crowdsourcing} collected relevance ratings for a small subset of audio samples and captions in Clotho~\cite{Drossos2020Clotho} via crowdsourced subjective assessments.
        Human annotators were asked to assign relevance ratings (ranging from 0 to 100) to indicate their judgements of how much the acoustic content of an audio sample matched with a given caption.
        Relevance ratings were collected for 17 audio samples per caption across 600 captions, resulting in a total of 10,200 ratings.
        We denoted this subset of Clotho with human-assigned relevance ratings as ``GrRel''.

        \textbf{Binary Relevances}.
        We constructed three datasets of audio samples and captions with binary relevances, where an audio sample was considered relevant to its corresponding caption but irrelevant to all other captions in the dataset.
        Specifically, we denoted as ``BiRel'' the set of audio samples and captions, the same as ``GrRel'', but annotated with binary relevances.
        The ``SuperBiRel'' consisted of audio samples in ``BiRel'' and ``GrRel'', each accompanied by one of the five reference captions provided in Clotho~\cite{Drossos2020Clotho}.
        Finally, all audio samples and captions in Clotho~\cite{Drossos2020Clotho} were utilized for experiment.
        Regardless of the graded and binary relevances in these datasets, we had ``GrRel'' $=$ ``BiRel'' $\subset$ ``SuperBiRel'' $\subset$ Clotho.
        Table~\ref{tab:experiment_dataset} summarizes the four datasets.

        \begin{table}[!t]
            \setlength{\tabcolsep}{3pt}
            \renewcommand{\arraystretch}{1.2}
            \centering
            \begin{tabular}{c|c|l|l|l}
                \hline
                \multirow{2}{*}{\bfseries Dataset} & \multirow{2}{*}{\bfseries Relevance} & \multicolumn{3}{c}{\bfseries \#Audio / \#Captions} \\
                \cline{3-5}
                &                         & \bfseries development & \bfseries validation & \bfseries evaluation \\
                \hline
                GrRel      & Graded                  & 2186 / 200            & 1004 / 200           & 1009 / 200           \\
                \hline
                BiRel      & \multirow{3}{*}{Binary} & 2186 / 200            & 1004 / 200           & 1009 / 200           \\
                \cline{1-1} \cline{3-5}
                SuperBiRel &                         & 2186 / 2186           & 1004 / 1004          & 1009 / 1009          \\
                \cline{1-1} \cline{3-5}
                Clotho     &                         & 3839 / 19195          & 1045 / 5225          & 1045 / 5225          \\
                \hline
            \end{tabular}
            \caption{Audio samples and captions with human-assigned relevance ratings (i.e., graded relevances) and binary relevances.}
            \label{tab:experiment_dataset}
            \vspace{-12pt}
        \end{table}

        \subsection{Audio and Text Encoders}\label{subsec:audio-and-text-encoders}

        \textbf{Audio Encoder}.
        A pretrained CNN14~\cite{Kong2020PANNs} was utilized as the audio encoder, with a fully-connected layer added on its top.
        It took 64-dimensional log mel-band energies as inputs, which were computed from 40 ms Hanning-windowed frames with a hop length of 20 ms, and generated 300-dimensional audio embeddings.
        We fine-tuned the audio encoder during training.

        \textbf{Text Encoder}.
        The Sentence-BERT (specifically, ``all-mpnet-base-v2'')~\cite{Reimers2019Sentence} was employed as the text encoder to extract 768-dimensional text embeddings from audio captions.
        An additional fully-connected layer was added on top to transform these text embeddings into 300-dimensional embeddings.
        The Sentence-BERT was frozen during training.

        \textbf{Training Setup}.
        During training, optimization was carried out using the Adam optimizer, starting with a learning rate of $0.001$.
        If the validation loss failed to improve over five consecutive epochs, the learning rate was reduced by a factor of ten.
        Early stopping was applied with a patience of ten epochs to terminate training if no improvement was observed.

        \subsection{Language-based Audio Retrieval}\label{subsec:language-based-audio-retrieval}

        The proposed method was validated on language-based audio retrieval, a downstream task of audio-text relevance learning, which aims to retrieve audio samples from a dataset based on their relevance to a given textual query.
        We performed language-based audio retrieval by using captions as textual queries to retrieve their corresponding audio samples in the Clotho evaluation set (``Clotho-evaluation'')~\cite{Drossos2020Clotho}.
        Audio-text relevance was measured by the cosine similarity between audio and text embeddings, with higher cosine similarity indicating greater relevance.

        \textbf{Evaluation Metrics}.
        Retrieval performance was assessed using mean Average Precision (mAP) and Recall at 10 (R@10), as done in~\cite{Xie2022Language}.
        The mAP was determined by averaging the Average Precision (AP) scores across all query captions, where AP was the average of the precision values at the positions of audio samples in the relevance-based ranked list corresponding to a query caption.
        R@10 was calculated as the proportion of audio samples within the top-10 results relative to the total number of audio samples corresponding to a query caption, averaged over all captions.
        Higher values for both metrics indicate better performance.

        \vspace{-6pt}

        \section{Results}\label{sec:results}
        \vspace{-6pt}

        This section reports the experimental results for language-based audio retrieval and audio-text relevance learning.

        \subsection{Language-based Audio Retrieval}\label{subsec:audio-retrieval}

        \begin{table}[!t]
            \setlength{\tabcolsep}{3pt}
            \renewcommand{\arraystretch}{1.2}
            \centering
            \begin{tabular}{c|c|c|c}
                \hline
                \multirow{2}{*}{\bfseries Training Dataset} & \multirow{2}{*}{\bfseries Loss} & \multicolumn{2}{c}{\bfseries Evaluation Metrics} \\
                \cline{3-4}
                &                          & \bfseries mAP     & \bfseries R@10    \\
                \hline
                GrRel              & \multirow{2}{*}{ListNet} & 0.034 $\pm$ 0.001 & 0.070 $\pm$ 0.002 \\
                \cline{1-1} \cline{3-4}
                BiRel              &                          & 0.015 $\pm$ 0.002 & 0.024 $\pm$ 0.004 \\
                \hline
                SuperBiRel         & \multirow{2}{*}{InfoNCE} & 0.168 $\pm$ 0.005 & 0.356 $\pm$ 0.010 \\
                \cline{1-1} \cline{3-4}
                Clotho             &                          & 0.239 $\pm$ 0.001 & 0.482 $\pm$ 0.002 \\
                \hline
                SuperBiRel + GrRel & \multirow{2}{*}{Joint}   & 0.173 $\pm$ 0.001 & 0.364 $\pm$ 0.009 \\
                \cline{1-1} \cline{3-4}
                Clotho + GrRel     &                          & 0.244 $\pm$ 0.002 & 0.486 $\pm$ 0.002 \\
                \hline
            \end{tabular}
            \caption{Language-based audio retrieval on Clotho-evaluation.}
            \label{tab:retrieval_results}
            \vspace{-12pt}
        \end{table}

        Table~\ref{tab:retrieval_results} presents the results on Clotho-evaluation.
        Each evaluation is repeated five times, and the averaged metrics are reported.

        \textbf{Continuous and Binary Relevances}.
        Note that the ListNet loss (see Section~\ref{subsec:learning-objectives}) can also work with binary relevances.
        In such a case, the binary relevance values $\left\{ 0, 1 \right\}$ are mapped to relevance ratings $\left\{ 0, 100 \right\}$, respectively.
        When training the dual-encoder framework with the ListNet loss~\eqref{eq:listnet_loss}, we experimented with the same audio samples and captions with either human-assigned relevance ratings (``GrRel'') or binary relevances (``BiRel'').
        Experimental results show that using human-assigned relevance ratings for training leads to better performance.
        For instance, ``GrRel'' surpasses ``BiRel'', achieving an R@10 score of $0.070 \pm 0.002$ compared to $0.024 \pm 0.004$ for ``BiRel''.
        We conclude that continuous relevances (e.g., human-assigned relevance ratings) outperform binary relevances in depicting the relevance between audio samples and texts, thereby resulting in superior performance in language-based audio retrieval.

        \textbf{Learning Objectives}.
        When training the dual-encoder framework with different learning objectives, the joint loss~\eqref{eq:joint_loss} achieves enhanced performance compared to both the ListNet loss~\eqref{eq:listnet_loss} and the InfoNCE loss~\cite{Oord2018Representation}.
        For instance, the InfoNCE loss with ``SuperBiRel'' yields an R@10 score of $0.356 \pm 0.010$, and the ListNet loss with ``GrRel'' obtains an R@10 score of $0.070 \pm 0.002$.
        When combined, the joint loss with ``GrRel'' and ``SuperBiRel'' attains an R@10 score of $0.364 \pm 0.009$.
        This demonstrates the effectiveness of the proposed method in utilizing both human-assigned relevance ratings and binary relevances for audio-text relevance learning.

        Additionally, regardless of the learning objectives, the volume of training data (e.g., the number of audio samples and captions) affects performance.
        For instance, when working with the InfoNCE loss, the larger Clotho outperforms ``SuperBiRel'', achieving an R@10 score of $0.482 \pm 0.002$ compared to $0.356 \pm 0.010$ for ``SuperBiRel''.
        The ListNet loss obtains the worst performance, likely due to the limited number of audio samples and captions in ``GrRel'' and ``BiRel''.

        \subsection{Learned Audio-Text Relevances}\label{subsec:learned-audio-text-relevances}

        The learned audio-text relevances are measured by the cosine similarity between audio and text embeddings, with higher similarity indicating greater relevance.
        We collect learned audio-text relevances for audio samples and captions in the ``GrRel'' evaluation set from three setups: the ListNet loss~\eqref{eq:listnet_loss} with ``GrRel'', the InfoNCE loss~\cite{Oord2018Representation} with ``SuperBiRel'', and the joint loss~\eqref{eq:joint_loss} with both datasets.
        These learned relevances are compared with human-assigned relevance ratings in the ``GrRel'' evaluation set, which includes 3,400 relevance ratings across 1,009 audio samples and 200 captions.
        Specifically, we calculate Spearman's rank-order correlation~\cite{Spearman1904} between the learned audio-text relevances and human-assigned relevance ratings.

        \begin{table}[!t]
            \setlength{\tabcolsep}{3pt}
            \renewcommand{\arraystretch}{1.2}
            \centering
            \begin{tabular}{c|c|c|c}
                \hline
                \multirow{2}{*}{\bfseries Training Dataset} & \multirow{2}{*}{\bfseries Loss} & \multicolumn{2}{c}{\bfseries Correlation} \\
                \cline{3-4}
                &         & \bfseries $\rho$-statistic & \bfseries p-value \\
                \hline
                GrRel              & ListNet & 0.530                      & $<$ 0.001         \\
                \hline
                SuperBiRel         & InfoNCE & 0.671                      & $<$ 0.001         \\
                \hline
                SuperBiRel + GrRel & Joint   & 0.690                      & $<$ 0.001         \\
                \hline
            \end{tabular}
            \caption{Spearman's rank-order correlation between learned relevances and human-assigned ratings in the ``GrRel'' evaluation set.}
            \label{tab:correlation_results}
        \end{table}

        As shown in Table~\ref{tab:correlation_results}, all three setups learn audio-text relevances that are moderately positively correlated with the human-assigned relevance ratings ($0.4<\rho<0.7$, p-values $<0.001$).
        Employing the joint loss with both datasets yields the highest correlation observed with the human-assigned relevance ratings ($\rho = 0.690$, p-value $< 0.001$).
        Despite the ListNet loss applied to ``GrRel'', which incorporates the fewest audio samples and captions during training, it demonstrates a moderate positive correlation ($\rho = 0.53$, p-value $< 0.001$), showing the effectiveness of the proposed method in audio-text relevance learning.

        \vspace{-6pt}

        \section{Analysis of crowdsourced ratings}\label{sec:complexity}
        \vspace{-6pt}

        \begin{table}[!t]
            \setlength{\tabcolsep}{3pt}
            \renewcommand{\arraystretch}{1.2}
            \centering
            \begin{tabular}{|p{0.25cm}|p{1.9cm}|p{1.2cm}|p{1.2cm}|p{1.2cm}|p{1.2cm}|p{1.2cm}|}
                \hline
                & \makecell[tl]{Feature}        & \makecell[tl]{HR} & \makecell[tl]{MR} & \makecell[tl]{D(H, M)} & \makecell[tl]{APT} \\ \hline\hline
                \multirow{3}{*}{\rotatebox{90}{Audio}} &
                \makecell[tl]{e-time} &
                    {n.s.} & {n.s.} & n.s. & n.s.  \\ \cline{2-6}

                & \makecell[tl]{e-class}  & -0.195**          & -0.233**          & n.s.                   & 0.089*             \\ \cline{2-6}

                & \makecell[tl]{audio duration} & {n.s.}            & {n.s.}            & {n.s.}                 & 0.604**            \\ \hline\hline
                \multirow{7}{*}{\rotatebox{90}{Text}}
                & perplexity                    & -0.092*           & n.s.              & n.s.                   & {n.s.}             \\ \cline{2-6}
                & \# words                      & 0.130**           & {n.s.}            & 0.108**                & {0.227**}          \\ \cline{2-6}
                & \# C-words                    & 0.099*            & n.s.              & {0.100*}               & 0.213**            \\ \cline{2-6}
                & \# nouns                      & {n.s.}            & {n.s.}            & {0.107**}              & 0.158**            \\ \cline{2-6}
                & \# adjectives                 & {n.s.}            & {n.s.}            & {n.s.}                 & 0.082*             \\ \cline{2-6}
                & \# fr-words                   & {0.095*}          & {n.s.}            & {n.s.}                 & 0.113**            \\ \cline{2-6}
                & \# fr-C-words                 & {0.089*}          & {n.s.}            & {n.s.}                 & 0.127**            \\  \cline{2-6}
                & \# fr-nouns                   & {n.s.}            & {n.s.}            & {0.083*}               & 0.095*             \\ \hline
            \end{tabular}
            \caption{ Pearson correlation coefficients between annotation characteristics and data features. * p-value \textless 0.05, ** p-value \textless 0.01, "n.s." = not significant.}
            \label{Tab-complexity}
            \vspace{-12pt}
        \end{table}

        To better understand the data-related bias factors in our audio-text relevance annotations and learning systems, we analyzed how text and audio properties might be related to the relevance ratings provided by humans and learned by the machine. Previous studies in crowd-sourcing sound events have linked annotation characteristics like inter-annotator agreement to audio attributes such as overlapping sounds \cite{martin2023strong} and sound-event loudness \cite{cartwright2018investigating}. Regarding data attributes, research has explored audio complexity through spectral dynamics \cite{singh2011measuring} and cognitive processing demands \cite{lang2015conceptualizing}, as well as text complexity based on caption length and syntactic structure \cite{brunato2018sentence}.

        Inspired by these works, we measured a set of audio and text features and analyzed their correlation with human and machine audio-caption relevance ratings. We conducted our analysis across true-positive (TP) pairs (i.e., 600 crowdsourced captions and their original audio pairs \cite{Drossos2020Clotho}, see \cite{Xie2023Crowdsourcing}).
        For audio clips, we utilized the probability matrix from the pre-trained sound event detection PANNS model \cite{Kong2020PANNs}, measuring entropy across 527 sound classes (e-class) using averaged probabilities across time, and entropy over time (e-time) using averaged probabilities across classes, where entropy \( H = -\sum_{i} p_i \log_{2}(p_i) \), with \( p_i \) as probability distributions. Additionally, we considered the audio clip duration as a third attribute. Regarding text captions, we analyzed perplexity (as a measure of syntactic complexity), word count, content words (C-words) count, and adjectives count. Additionally, we compiled lists of the 500 most frequent (fr) words, content words, and adjectives from the original Clotho dataset, tallying their occurrences in each caption.

        Audio and text attributes were correlated against the human-rated (HR) audio-text relevances and their standard deviation across annotators (SD-HR), the latter being a measure of annotators' disagreement on the relevances. The attributes were also compared against machine relevance ratings (MR) from a model trained on Clotho data (InfoNCE loss with binary relevances) to understand the factors contributing to the learnability of the data. Finally, we explored whether the audio/text attributes can explain the degree of disagreement between HR and MR (D(H, M)), and whether the average time annotators spent playing audio clips (APT) correlates with audio/text attributes.

        Table \ref{Tab-complexity} presents the Pearson correlation coefficients between the measured audio and text features and the annotation characteristics. HR and MR are negatively correlated with class entropy, indicating that the presence of diverse sound classes in the clip leads both humans and machines to score a TP audio-text pair as less relevant. Conversely, a positive correlation (r=0.213, p\textless0.01) between std-HR and class entropy suggests that such a feature results in disagreement among annotators. Moreover, the syntactic complexity of captions tends to lead annotators to score a true positive audio-caption pair as less relevant. Conversely, longer and denser captions tend to lead annotators to perceive audio-caption pairs as more relevant. Similarly, the disagreement between human and machine ratings increases with longer and denser captions. As expected, annotators tended to play audio clips for a longer duration when the actual length of the audio was longer. Surprisingly, the average played time is also correlated with text attributes, suggesting that when annotators are presented with a longer and denser caption, they tend to listen to more of the audio clip before rating the relevance of the audio-caption pair.

        \vspace{-6pt}

        \section{Conclusions}\label{sec:conclusions}
        \vspace{-6pt}

        This study introduced an approach to audio-text relevance learning that integrated both continuous and binary relevances.
        By training modality-specific encoders, we projected audio samples and texts into a shared embedding space, where the cosine similarity of their embeddings served as a measure of their relevance.
        Through a combined optimization of a listwise ranking objective using continuous relevance ratings and a contrastive learning objective with binary relevances during training, our method demonstrated enhanced performance in language-based audio retrieval, a downstream task in this domain.
        Moreover, we analyzed how various properties of captions and audio clips influenced both human-assigned and machine-learned continuous relevances.

        \bibliographystyle{IEEEtran}
        \bibliography{ms}

\begin{thebibliography}{10}
\providecommand{\url}[1]{#1}
\def\UrlFont{\rmfamily}
\providecommand{\newblock}{\relax}
\providecommand{\bibinfo}[2]{#2}
\providecommand\BIBentrySTDinterwordspacing{\spaceskip=0pt\relax}
\providecommand\BIBentryALTinterwordstretchfactor{4}
\providecommand\BIBentryALTinterwordspacing{\spaceskip=\fontdimen2\font plus
\BIBentryALTinterwordstretchfactor\fontdimen3\font minus
  \fontdimen4\font\relax}
\providecommand\BIBforeignlanguage[2]{{%
\expandafter\ifx\csname l@#1\endcsname\relax
\typeout{** WARNING: IEEEtran.bst: No hyphenation pattern has been}%
\typeout{** loaded for the language `#1'. Using the pattern for}%
\typeout{** the default language instead.}%
\else
\language=\csname l@#1\endcsname
\fi
#2}}

\bibitem{Xie2022Language}
H.~Xie, S.~Lipping, and T.~Virtanen, ``Language-based audio retrieval task in
  dcase 2022 challenge,'' in \emph{Proc. Detect. Classif. Acoust. Scenes Events
  Work. (DCASE)}, 2022, pp. 216--220.

\bibitem{Wu2023Large}
Y.~Wu, K.~Chen, T.~Zhang, Y.~Hui, T.~Berg-Kirkpatrick, and S.~Dubnov,
  ``Large-scale contrastive language-audio pretraining with feature fusion and
  keyword-to-caption augmentation,'' in \emph{Proc. Int. Conf. Acoustic.,
  Speech and Signal Process. (ICASSP)}, 2023, pp. 1--5.

\bibitem{Primus2023Advancing}
P.~Primus, K.~Koutini, and G.~Widmer, ``Advancing natural-language based audio
  retrieval with passt and large audio-caption data sets,'' in \emph{Proc.
  Detect. Classif. Acoust. Scenes Events Work. (DCASE)}, 2023, pp. 151--155.

\bibitem{Drossos2020Clotho}
K.~Drossos, S.~Lipping, and T.~Virtanen, ``Clotho: An audio captioning
  dataset,'' in \emph{Proc. Int. Conf. Acoustic., Speech and Signal Process.
  (ICASSP)}, 2020, pp. 736--740.

\bibitem{Mei2023WavCaps}
X.~Mei, C.~Meng, H.~Liu, Q.~Kong, T.~Ko, C.~Zhao, M.~D. Plumbley, Y.~Zou, and
  W.~Wang, ``Wavcaps: A chatgpt-assisted weakly-labelled audio captioning
  dataset for audio-language multimodal research,'' \emph{IEEE/ACM Trans. Audio
  Speech Lang. Process.}, pp. 1--15, 2024.

\bibitem{Oord2018Representation}
A.~Van~den Oord, Y.~Li, and O.~Vinyals, ``Representation learning with
  contrastive predictive coding,'' 2018,
  \href{http://arxiv.org/abs/1807.03748}{arXiv preprint arXiv:1807.03748}.

\bibitem{Sakai2021Graded}
T.~Sakai, ``Graded relevance,'' in \emph{Evaluating Information Retrieval and
  Access Tasks: NTCIR's Legacy of Research Impact}.\hskip 1em plus 0.5em minus
  0.4em\relax Singapore: Springer Singapore, 2021, pp. 1--20.

\bibitem{Roitero2021On}
K.~Roitero, E.~Maddalena, S.~Mizzaro, and F.~Scholer, ``On the effect of
  relevance scales in crowdsourcing relevance assessments for information
  retrieval evaluation,'' \emph{Inf. Process. Manag.}, vol.~58, no.~6, p.
  102688, 2021.

\bibitem{Xie2023Crowdsourcing}
H.~Xie, K.~Khorrami, O.~Räsänen, and T.~Virtanen, ``Crowdsourcing and
  evaluating text-based audio retrieval relevances,'' in \emph{Proc. Detect.
  Classif. Acoust. Scenes Events Work. (DCASE)}, 2023, pp. 226--230.

\bibitem{Cao2007Learning}
Z.~Cao, T.~Qin, T.~Liu, M.~Tsai, and H.~Li, ``Learning to rank: from pairwise
  approach to listwise approach,'' in \emph{Proc. Int. Conf. Mach. Learn.
  (ICML)}, 2007, pp. 129--136.

\bibitem{Kong2020PANNs}
Q.~Kong, Y.~Cao, T.~Iqbal, Y.~Wang, W.~Wang, and M.~Plumbley, ``Panns:
  Large-scale pretrained audio neural networks for audio pattern recognition,''
  \emph{IEEE/ACM Trans. Audio Speech Lang. Process.}, pp. 2880--2894, 2020.

\bibitem{Reimers2019Sentence}
N.~Reimers and I.~Gurevych, ``Sentence-{BERT}: Sentence embeddings using
  siamese {BERT}-networks,'' in \emph{Proc. Empherical Methods Nat. Lang.
  Process. (EMNLP)}, 2019, pp. 3982--3992.

\bibitem{Spearman1904}
C.~Spearman, ``The proof and measurement of association between two things,''
  \emph{Am. J. Psychol.}, vol.~15, no.~1, pp. 72--101, 1904.

\bibitem{martin2023strong}
I.~Mart{\'\i}n-Morat{\'o} and A.~Mesaros, ``Strong labeling of sound events
  using crowdsourced weak labels and annotator competence estimation,''
  \emph{IEEE/ACM Trans. Audio Speech Lang. Process.}, vol.~31, pp. 902--914,
  2023.

\bibitem{cartwright2018investigating}
M.~Cartwright, J.~Salamon, A.~Seals, O.~Nov, and J.~P. Bello, ``Investigating
  the effect of sound-event loudness on crowdsourced audio annotations,'' in
  \emph{Proc. IEEE Int. Conf. Acoust. Speech Signal Process. (ICASSP)}, 2018,
  pp. 341--345.

\bibitem{singh2011measuring}
N.~C. Singh, ``Measuring the ‘complexity’of sound,'' \emph{Pramana},
  vol.~77, pp. 811--816, 2011.

\bibitem{lang2015conceptualizing}
A.~Lang, Y.~Gao, R.~F. Potter, S.~Lee, B.~Park, and R.~L. Bailey,
  ``Conceptualizing audio message complexity as available processing
  resources,'' \emph{Commun. Res.}, vol.~42, no.~6, pp. 759--778, 2015.

\bibitem{brunato2018sentence}
D.~Brunato, L.~De~Mattei, F.~Dell'Orletta, B.~Iavarone, G.~Venturi,
  \emph{et~al.}, ``Is this sentence difficult? do you agree?'' in \emph{Proc.
  Empir. Methods Nat. Lang. Process. (EMNLP)}, 2018, pp. 2690--2699.

\end{thebibliography}

    \end{sloppy}
\end{document}